\documentstyle[preprint,aps,epsf,floats]{revtex}

\begin{document}
\tighten
\def\si{{}^1\kern-.14em S_0}
\def\siii{{}^3\kern-.14em S_1}
\def\piii{{}^3\kern-.14em P_1}
\def\diii{{}^3\kern-.14em D_1}
\def\pone{{}^3\kern-.14em P_1}
\def\pzero{{}^3\kern-.14em P_0}
\def\ptwo{{}^3\kern-.14em P_2}
\newcommand{\gsim}{\raisebox{-0.7ex}{$\stackrel{\textstyle >}{\sim}$ }}
\newcommand{\lsim}{\raisebox{-0.7ex}{$\stackrel{\textstyle <}{\sim}$ }}
\def\pislash{ {\pi\hskip-0.6em /} }
\def\pislashsmall{ {\pi\hskip-0.375em /} }
\def\pslash{p\hskip-0.45em /}
\def\nopi{ {\rm EFT}(\pislash) }
\def\nopit{ {\rm dEFT}(\pislash) }
\def\Ltwo{ {^\pislashsmall \hskip -0.2em L_2 }}
\def\Lone{ {^\pislashsmall \hskip -0.2em L_1 }}
\def\CQuad{ {^\pislashsmall \hskip -0.2em C_{\cal Q} }}
\def\Czero{ {^\pislashsmall \hskip -0.2em C_{0}^{(\siii)} }}
\def\Czeromone{ {^\pislashsmall \hskip -0.2em C_{0,-1}^{(\siii)} }}
\def\Czerozero{ {^\pislashsmall \hskip -0.2em C_{0,0}^{(\siii)} }}
\def\Czeroone{ {^\pislashsmall \hskip -0.2em C_{0,1}^{(\siii)} }}
\def\Ctwo{ {^\pislashsmall \hskip -0.2em C_{2}^{(\siii)} }}
\def\Ctwomtwo{ {^\pislashsmall \hskip -0.2em C_{2,-2}^{(\siii)} }}
\def\Ctwomone{ {^\pislashsmall \hskip -0.2em C_{2,-1}^{(\siii)} }}
\def\Cfour{ {^\pislashsmall \hskip -0.2em C_{4}^{(\siii)} }}
\def\CSDzero{ {^\pislashsmall \hskip -0.2em C_0^{(sd)} }}
\def\CSDtwotwotwo{ {^\pislashsmall \hskip -0.2em C_{2,-2}^{(sd)} }}
\def\CSDzeromone{ {^\pislashsmall \hskip -0.2em C_{0,-1}^{(sd)} }}
\def\CSDzerozero{ {^\pislashsmall \hskip -0.2em C_{0,0}^{(sd)} }}
\def\CSDtwoone{ {^\pislashsmall \hskip -0.2em \tilde C_2^{(sd)} }}
\def\CSDtwotwo{ {^\pislashsmall \hskip -0.2em C_2^{(sd)} }}
\def\CSDzerotwo{ {^\pislashsmall \hskip -0.2em C_{0,0}^{(sd)} }}
\def\LX{ {^\pislashsmall \hskip -0.2em L_X }}
\def\CSDfour{ {^\pislashsmall \hskip -0.2em C_4^{(sd)} }}
\def\CSDfourt{ {^\pislashsmall \hskip -0.2em \tilde C_4^{(sd)} }}
\def\CSDfourtt{ {^\pislashsmall \hskip -0.2em {\tilde{\tilde C}}_4^{(sd)} }}
\def\etasd{\eta_{sd} }
\def\ZCzeromone{ 
{_z \hskip -0.4em {^\pislashsmall \hskip -0.2em C_{0,-1}^{(\siii)} }}}
\def\ZCzerozero{ 
{_z \hskip -0.4em {^\pislashsmall \hskip -0.2em C_{0,0}^{(\siii)} }}}
\def\ZCzeroone{ 
{_z \hskip -0.4em {^\pislashsmall \hskip -0.2em C_{0,1}^{(\siii)} }}}
\def\ZCtwomtwo{ 
{_z \hskip -0.4em {^\pislashsmall \hskip -0.2em C_{2,-2}^{(\siii)} }}}
\def\ZCtwomone{ 
{_z \hskip -0.4em {^\pislashsmall \hskip -0.2em C_{2,-1}^{(\siii)} }}}
\def\ZCfourmthree{ 
{_z \hskip -0.4em {^\pislashsmall \hskip -0.2em C_{4,-3}^{(\siii)} }}}
\def\rCzeromone{ 
{_\rho \hskip -0.4em {^\pislashsmall \hskip -0.2em C_{0,-1}^{(\siii)} }}}
\def\rCzerozero{ 
{_\rho \hskip -0.4em {^\pislashsmall \hskip -0.2em C_{0,0}^{(\siii)} }}}
\def\rCzeroone{ 
{_\rho \hskip -0.4em {^\pislashsmall \hskip -0.2em C_{0,1}^{(\siii)} }}}
\def\rCtwomtwo{ 
{_\rho \hskip -0.4em {^\pislashsmall \hskip -0.2em C_{2,-2}^{(\siii)} }}}
\def\rCtwomone{ 
{_\rho \hskip -0.4em {^\pislashsmall \hskip -0.2em C_{2,-1}^{(\siii)} }}}
\def\rCfourmthree{ 
{_\rho \hskip -0.4em {^\pislashsmall \hskip -0.2em C_{4,-3}^{(\siii)} }}}
\def\CPzero{ {^\pislashsmall \hskip -0.2em C^{(\pzero)}_2  }}
\def\CPone{ {^\pislashsmall \hskip -0.2em C^{(\pone)}_2  }}
\def\CPtwo{ {^\pislashsmall \hskip -0.2em C^{(\ptwo)}_2  }}

\def\Journal#1#2#3#4{{#1} {\bf #2}, #3 (#4)}

\def\NCA{\em Nuovo Cimento}
\def\NIM{\em Nucl. Instrum. Methods}
\def\NIMA{{\em Nucl. Instrum. Methods} A}
\def\NPB{{\em Nucl. Phys.} B}
\def\NPA{{\em Nucl. Phys.} A}
\def\NP{{\em Nucl. Phys.} }
\def\PLB{{\em Phys. Lett.} B}
\def\PRL{\em Phys. Rev. Lett.}
\def\PRD{{\em Phys. Rev.} D}
\def\PRC{{\em Phys. Rev.} C}
\def\PRA{{\em Phys. Rev.} A}
\def\PR{{\em Phys. Rev.} }
\def\ZPC{{\em Z. Phys.} C}
\def\SJP{{\em Sov. Phys. JETP}}
\def\SJNP{{\em Sov. Phys. Nucl. Phys.}}

\def\FBS{{\em Few Body Systems Suppl.}}
\def\IJMP{{\em Int. J. Mod. Phys.} A}
\def\UJP{{\em Ukr. J. of Phys.}}
\def\CJP{{\em Can. J. Phys.}}
\def\SCI{{\em Science} }
\def\AST{{\em Astrophys. Jour.} }
\def\tran{dibaryon}
\def\trans{dibaryons}
\def\Tran{Dibaryon}
\def\Trans{Dibaryons}
\def\TRANS{DIBARYONS}
\def\yt{y}

\def\hpi{ h_{\pi NN}^{(1)} }
\def\hwk{ h_{33}^{(1)} }

\preprint{\vbox{
\hbox{ NT@UW-00-035}
}}
\bigskip
\bigskip

\title{Parity Violation in Low-Energy
$np\rightarrow d\gamma$ and the Deuteron Anapole Moment
}

\author{{\bf Martin J. Savage}$^{a,b}$}
\address{$^a$ Department of Physics, University of Washington, \\
Seattle, WA 98195. }
\address{$^b$ Jefferson Laboratory, 12000 Jefferson Avenue,\\
Newport News, VA 23606.}
\maketitle

\begin{abstract}
Parity violation in low-energy nuclear observables
is included in the pionless effective field theory.  
The  model-independent relation between the 
parity-violating asymmetry in
$\vec np\rightarrow d\gamma$ and the non-nucleon part of the 
deuteron anapole moment is discussed.
The asymmetry in
$\vec np\rightarrow d\gamma$ computed with KSW 
power-counting, and recently criticized by Desplanques, is discussed.
\end{abstract}

\vskip 2in

\leftline{December 2000}
\vfill\eject

\section{Introduction}

Parity violation continues to be a focus 
of the nuclear physics community.
It is being used both to uncover the structure of the nucleon in 
electron-scattering 
experiments such as SAMPLE\cite{sample},
and to determine parity-violating (PV)
but flavour-conserving couplings between
pions and nucleons\cite{cesium,lightN}.
While the PV interaction between leptons and nucleons 
is well understood, 
the purely hadronic PV sector is presently somewhat confused.
Recently it has been pointed out that measurements 
of PV observables in the  single nucleon sector would significantly
improve this situation~\cite{BSpv,Chpv}

The conventional description of nuclear parity-violation is formulated 
in terms 
of a PV potential generated through single-meson exchange (SME), 
such as the $\pi$, $\rho$ and $\omega$\cite{DDH}.
The experimental measurements of parity-violation in light nuclei\cite{lightN}
do not yet indicate that the SME picture is consistent.
In addition, in the framework of SME,
the data from $^{18}F$ suggest that the  $\Delta I=1$
pion-nucleon weak coupling constant, $\hpi$,
is much smaller than naive dimensional analysis (NDA) would suggest\cite{DDH}.
If, in fact, $\hpi$ is much smaller than the NDA estimate 
then weak one-pion-exchange (OPE) may not
provide the dominant long-distance component of the PV potential.
It becomes necessary to include parity-violating 
two-pion~\cite{twopi,KS}
and multi-pion interactions
consistent with the chiral symmetries of QCD as detailed in
Ref.~\cite{KS} and further developed in Ref.~\cite{ZPHM}.
For very low-energy PV observables, the underlying mechanism
generating parity-violation does not need to be known in order to make
model-independent relations between different observables.
This approach has been applied to PV in the NN-sector for many 
years~\cite{Da65,Mi76,DeMi78} and we continue to develop
this approach using the pionless
effective field theory defined with dibaryon fields~\cite{Ka97},
$\nopit$~\cite{BS00}.
The difference between $\nopit$ and the more familiar $\nopi$~\cite{Ch99}
is that effective range contributions are summed to all orders with 
$\nopit$~\footnote{
The motivation for performing this re-ordering of the perturbative expansion
is the relatively large size of the product 
$\gamma r^{(\siii)}\sim 0.4$, where $\gamma$ is the deuteron binding momentum 
and $ r^{(\siii)}$ is the effective range in the $\siii$ channel.
For processes involving the deuteron, the convergence is 
substantially improved by resumming such terms, as the normalization of the 
deuteron s-state plays a central role. 
Emperically, it has been found that these terms make the largest contribution
at each order in the $\nopi$ expansion.
After resumming such terms, 
the remaining contributions from the NN scattering amplitude 
involve the shape-parameter $r_1$ and higher order 
terms in the effective range expansion.
}.

Low-energy PV observables such as the 
deuteron anapole moment and the forward-backward asymmetry,
$A_\gamma$,
in polarized neutron capture $\vec n p\rightarrow d\gamma$,
can be described by an effective field theory (EFT)
involving only  nucleons and photons~\cite{Ch99}.
The strong interactions between nucleons in an s-wave are most economically
described by dibaryon fields~\cite{Ka97,BS00}, with a Lagrange density
\begin{eqnarray}
{\cal L}_t & = & 
N^\dagger \left[ i\partial_0 + {\nabla^2\over 2 M_N}\right] N
\ -\ 
t^{\dagger}_a \left[  i\partial_0 + {\nabla^2\over 4 M_N} - \Delta \right] t^a
\ -\  \yt \left[\  t^\dagger_j \ N^T P^j N\ +\ {\rm h.c.} \right]
\ \ \ ,
\label{eq:trandef}
\end{eqnarray}
where $t_i$ is the $\siii$-dibaryon,
$P^j$ is the spin-isospin projector for the $\siii$ channel, and 
$y$ is the coupling between  $t_i$ and two
nucleons in the $\siii$ channel.
A similar Lagrange density exists for nucleons in the $\si$-channel.
%
\begin{figure}[!ht]
\centerline{{\epsfxsize=6in \epsfbox{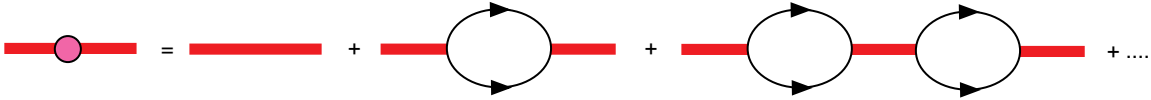}} }
\vskip 0.15in
\noindent
\caption{\it The dressed dibaryon propagator. 
The bare dibaryon propagator
is dressed by nucleon bubbles to all orders. 
}
\label{fig:trans}
\vskip .2in
\end{figure}
This Lagrange density alone 
reproduces the NN scattering amplitude 
when the class of diagrams shown in fig~\ref{fig:trans}
are summed,
with
\begin{eqnarray}
\yt^2 & = & {8\pi\over M_N^2 r^{(\siii)}}
\ \ ,\ \ 
\Delta\ =\ {2\over M_N  r^{(\siii)}} \left({1\over a^{(\siii)}} - \mu\right)
\ \ \ ,
\label{eq:coupfix}
\end{eqnarray}
where $\mu$ is the renormalization scale. 
A complete discussion of the power-counting and implementation of the 
dibaryon fields in the two-nucleon sector can be found in 
Refs.~\cite{Ka97,BS00}.
Weak interactions will be described by a Lagrange density of four-nucleon
operators in $\nopi$, which are written in terms of s-wave dibaryon fields
in $\nopit$.
For both the deuteron anapole moment and $A_\gamma$ we need only consider
PV interactions of the $\siii$ channel.
Restricting ourselves to $\Delta I=1$ parity-violation means that we consider 
only a $\siii-\piii$ coupling, described by a Lagrange density of the form
\begin{eqnarray}
{\cal L}_{wk} & = & 
i {h_{33}^{(1)}\over\sqrt{8 M_N r^{(\siii)}}}
\epsilon^{ijk}\ 
t^\dagger_i \ N^T \ \sigma_2\sigma_j\ \tau_2\tau_3\ 
{1\over 2} \left(i\overleftarrow\nabla - i\overrightarrow\nabla\right)_k\ N
\ +\ {\rm h.c.}
\ +\ ...
\ \ \ ,
\label{eq:lagwk}
\end{eqnarray}
where $h_{33}^{(1)}$ is an unknown weak coupling constant that must be 
fit to data or predicted from the standard model of electroweak interactions.
The ellipses denote higher dimension operators involving more powers of the 
center of mass energy, insertions of the electromagnetic gauge field
or isospin breaking effects, each suppressed by inverse powers of 
the pion mass (or the appropriate high scale).

A similar approach to PV interactions has been recently explored by 
Khriplovich and Korkin~\cite{KKnp}, where they use effective short range
interactions to describe the asymmetry in $\vec\gamma d\rightarrow np$
between the photon circular polarization states.
This process depends only upon the $\Delta I =0, 2$ PV interactions, and does
not receive a contribution from OPE in the SME description.


\section{Asymmetry in $\vec n p\rightarrow d\gamma$}

The forward-backward asymmetry in $\vec n p\rightarrow d\gamma$ is
defined by the coefficient $A_\gamma$ in
\begin{eqnarray}
{1\over\Gamma} {d\Gamma\over d\cos\theta} 
& = & 1 + A_\gamma\ \cos\theta
\ \ \ ,
\label{eq:asymm}
\end{eqnarray}
where $\Gamma$ is the width for $np\rightarrow d\gamma$,
and
where $\theta$ is the angle between the emitted photon momentum and the 
polarization direction of the incident neutron.
An upcoming measurement~\cite{snow} of $A_\gamma$ will determine the 
coupling $h_{33}^{(1)}$ appearing in eq.~(\ref{eq:lagwk}).
The asymmetry is generated by an interference between neutron capture from
the $\si$ channel through a strong interaction emitting an M1 photon, and 
capture from the $\siii$ channel via the weak interaction emitting an E1
photon.
The matrix element for $np\rightarrow d\gamma$ at threshold 
is~\cite{KSSWa}
(retaining only the parity conserving $M1$ amplitude $Y$
and the PV $E1$ amplitude $W$) 
\begin{eqnarray}
{\cal M} & = & 
i \ e\  Y \ \epsilon_{ijk} \ {\bf \epsilon}_{(d)}^{*,i} \ {\bf k}^j\  
{\bf \epsilon}_{(\gamma)}^{*,k} \ 
N^T\tau_2\tau_3\sigma_2 N
\ +\ 
i \ e\  W \ \epsilon_{ijk} \ {\bf \epsilon}_{(d)}^{*,i} \  
{\bf \epsilon}_{(\gamma)}^{*,k} \ 
N^T\tau_2\sigma_2\sigma^j N
\ \ \ ,
\label{eq:npamp}
\end{eqnarray}
where ${\bf k}$ is the outgoing photon three-momentum and $N$ is a 
nucleon iso-spinor.
In terms of $Y$ and $W$, the PV asymmetry is~\cite{KSSWa}
\begin{eqnarray}
A_\gamma & = & 
-{2 M_N\over\gamma^2} {{\rm Re}\left[ Y^* W\right]\over |Y|^2}
\ \ \ ,
\label{eq:asymmdef}
\end{eqnarray}
where $\gamma=\sqrt{M_N B}$ is the deuteron binding momentum, with $B$ the
deuteron binding energy.
The amplitudes $Y$ and $W$ are
\begin{eqnarray}
Y & = & 
{1\over M_N} \sqrt{\pi\over\gamma^3} {1\over\sqrt{1-\gamma r^{(\siii)}}}
\left[\  \kappa_1\ \left(1-\gamma a^{(\si)}\right)
\ -\ {1\over 2}\gamma^2 a^{(\si)} L_1\ \right]
\nonumber\\
W & = & 
{ h_{33}^{(1)} \over 4\sqrt{2}} \sqrt{\gamma\over M_N}
{1\over\sqrt{1-\gamma r^{(\siii)}}}
\left( 1 - {\gamma a^{(\siii)}\over 3}\right)
\ \ \ ,
\label{eq:swamps}
\end{eqnarray}
where $\kappa_1$ is the isovector magnetic moment of the nucleon.
These expressions for $W$, and $Y$ with $L_1=0$ are directly related to 
amplitudes determined in previous zero-range interaction computations,
e.g.~\cite{Da65,LaMC76}.
We have retained the formally higher order contribution 
from the gauge-invariant four-nucleon-one-photon interaction 
proportional to $L_1$, that couples the $\siii$ and $\si$
dibaryons and the magnetic field, described by the Lagrange density
\begin{eqnarray}
{\cal L}^{\bf B} & = & 
{e\over 2 M_N} N^\dagger \left( \kappa_0+\kappa_1 \tau_3\right)
\sigma\cdot {\bf B}\  N
\ +\ 
e\ { L_1\over M_N \sqrt{ r^{(\si)}\ r^{(\siii)}}}
\ t^{j\dagger} s_3 {\bf B}_j
\ \ \ +\ \  {\rm h.c.}
\ \ \ ,
\end{eqnarray}
where $s_a$ is the $\si$ dibaryon.
This interaction has not been included in previous zero-range
computations, with the exception of those recently performed 
with EFT's.
The well measured cross section for cold neutron capture 
$np\rightarrow d\gamma$ fixes $L_1$~\cite{BS00,Ch99,KSSWa}.

The determination of $\hwk$ from this process 
is model independent and can then be used to compute other
low-energy PV observables that depend only on this channel (otherwise
additional information will be required).
Corrections to this result are suppressed by 
powers of $Q\sim \gamma/m_\pi \sim 0.3$, where
one particular contribution is of the form
\begin{eqnarray}
{\cal L}^{\bf E} & = & 
i e\ { R_1\over M_N \sqrt{ r^{(\si)}\ r^{(\siii)}}}
\ \epsilon_{ijk}\ t^{i\dagger} t^j {\bf E}^k
\ \ \ +\ \  {\rm h.c.}
\ \ \ ,
\label{eq:wkfour}
\end{eqnarray}
that has not been included in previous PV discussions.
${\bf E}_j$ is the electric field operator  and 
$R_1$ is an unknown counterterm that must be determined
experimentally.


\section{Deuteron Anapole Form Factor}

\begin{figure}[!ht]
\centerline{{\epsfxsize=2in \epsfbox{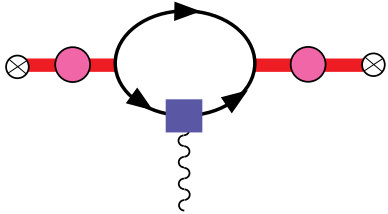}} }
\vskip 0.15in
\noindent
\caption{\it 
Single nucleon contributions to the deuteron anapole 
moment.
The solid square denotes an insertion of the single nucleon
anapole operator.
}
\label{fig:SN}
\vskip .2in
\end{figure}

\begin{figure}[!ht]
\centerline{{\epsfxsize=4in \epsfbox{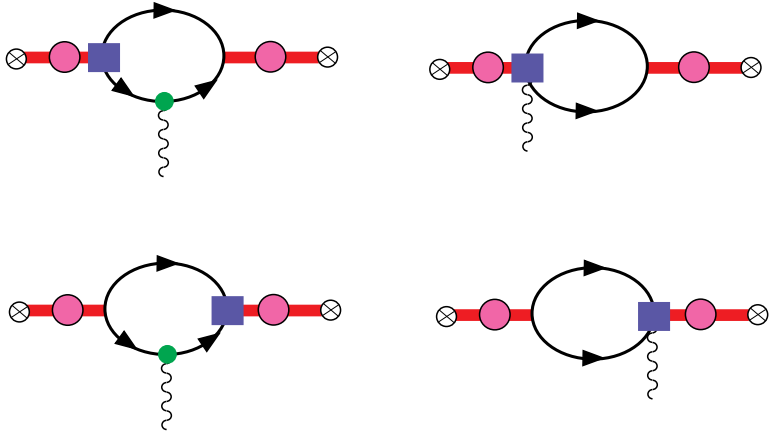}} }
\vskip 0.15in
\noindent
\caption{\it 
Non-nucleon contributions to the 
deuteron anapole moment.
The solid square denotes an insertion of the 
parity violating $\siii-\piii$ coupling, 
while the small solid circle denotes an insertion
of the nucleon magnetic moment or a minimally 
coupled interaction.
}
\label{fig:deut}
\vskip .2in
\end{figure}

The anapole moment is the PV coupling between a particle with non-zero spin and
the electromagnetic field.
As the operator vanishes for on-shell photons, it manifests itself as local
interactions between the particle and electrically charged particles.
For the deuteron, the anapole interaction is described by a 
Lagrange density of  the form
\begin{eqnarray}
{\cal L} & = & 
i A_d {1\over M_N^2} \ 
\epsilon_{ijk} \ d^{\dagger i} \ d^j \partial_\mu \ F^{\mu k}
\ \ \ ,
\end{eqnarray}
where $d^j$ is the deuteron annihilation operator.
$A_d$ receives two contributions, one from the anapole interactions of the
nucleon $A_N$, and one from the PV interaction between nucleons,
$A_h$, defined by $h_{33}^{(1)}$.
The single nucleon contribution, is defined to be $A_N$~\cite{SSa,SSb}, 
\begin{eqnarray}
A_N & = & 
\left( A_n\ +\ A_p\right)
\ {1\over 1-\gamma r^{(\siii)}}\ 
{4\gamma\over |{\bf k}|}
\tan^{-1}\left({ |{\bf k}|\over 4\gamma}\right)
\ \ \ ,
\label{eq:anaN}
\end{eqnarray}
arising from diagrams of the form shown in fig.~\ref{fig:SN},
%
where $A_{n,p}$ are the neutron and proton anapole moments, respectively.
If the weak one-pion nucleon interaction dominates the long-distance 
PV interaction, then at leading order in the chiral
expansion~\cite{HolMus}
\begin{eqnarray}
A_n & = & A_p \ =\ 
-{e g_A h_{\pi NN}^{(1)} M_N^2\over 48\pi f_\pi m_\pi}
\ \ \ ,
\label{eq:anaSN}
\end{eqnarray}
in the low-energy regime~\cite{SSa,SSb,HolMus,vKM}.
$f_\pi$ is the pion decay constant.
For the purposes of this discussion it is $A_h$ that is more interesting.
The nucleon magnetic moment and the minimally coupled electric 
interaction give  rise to a contribution of the form
\begin{eqnarray}
A_h \ =\ 
-{e h_{33}^{(1)} M_N^{3/2}\over 8\sqrt{2\pi}}
\ {1\over 1-\gamma r^{(\siii)}}\ 
& & 
\left(
\kappa_1 
{4\gamma\over |{\bf k}|}
\tan^{-1}\left({ |{\bf k}|\over 4\gamma}\right)
\right.\nonumber\\
& & \left.
\ +\ 
{4\over |{\bf k}|^2} \left[ \gamma^2 - 
\left(\gamma^2 + {|{\bf k}|^2\over 16}\right) 
{4\gamma\over |{\bf k}|}
\tan^{-1}\left({ |{\bf k}|\over 4\gamma}\right)
\right]
\right)
\ \ \ ,
\end{eqnarray}
due to the diagrams shown in fig.~\ref{fig:deut}~\footnote{
I expect, though have not checked, that this result can be obtained
from Ref.~\cite{HeHw82} when effective range theory 
wavefunctions are inserted into eq.(2) of Ref.~\cite{HeHw82}.
}.
%
In the zero-momentum transfer limit $A_h$ reduces to 
\begin{eqnarray}
A_h &\rightarrow &
-{e h_{33}^{(1)} M_N^{3/ 2}\over 8\sqrt{2\pi}}
\ {1\over 1-\gamma r^{(\siii)}}\ 
\left( \kappa_1 - {1\over 6} \right)
\ \ \ ,
\label{eq:analim}
\end{eqnarray}
consistent with limiting values determined 
previously~\cite{SSa,SSb,KKana}.

Therefore, it is clear that there is a model-independent relation
between $A_\gamma$ and the non-nucleon contribution to the deuteron anapole
moment at leading order in the effective field theory expansion.
In the theory with perturbative pions and KSW 
power-counting~\cite{Ka98}, 
both $A_\gamma$~\cite{KSSWa}
and the deuteron anapole form factor~\cite{SSa,SSb}
have been computed at leading order in $\hpi$
and  this relation  is found to exist.
However, we see that it is more general than leading order in $\hpi$.
Corrections to this relation arise from not only operators involving 
insertions of time-derivatives in PV operators
but also from insertions of the shape parameter
(and higher NN interactions terms) and from the anapole moment of the dibaryon 
itself,
\begin{eqnarray}
{\cal L} & = & 
i A_t {1\over M_N^3 r^{(\siii)}} \ 
\epsilon_{ijk} \ t^{\dagger i} \ t^j \partial_\mu \ F^{\mu k}
\ \ \ .
\end{eqnarray}
This gives a contribution to $A_d$ of
\begin{eqnarray}
\delta A_d & = & 
{\gamma\over M_N (1-\gamma r^{(\siii)})}\  A_t
\ \ \ ,
\end{eqnarray}
suppressed by just one power of $Q$ in the power-counting.
Thus, the relation between $A_\gamma$ and $A_d$ is expected to 
hold only up to order $Q$.


\section{Signs, EFT and All That}

It is appropriate to address some concerns recently expressed
about EFT calculations of PV processes~\cite{Des}.
In Ref.~\cite{Des} much was made of the difference in both the sign and 
magnitude of the calculation of $A_\gamma$ in Ref.~\cite{KSSWa} using 
an EFT with perturbative pions and KSW power-counting.
The relative sign between the PV asymmetry $A_\gamma$ arising from 
weak OPE alone and the PV OPE NN potential
computed in Ref.~\cite{KSSWa}
agrees with previous calculations, as detailed in Ref.~\cite{Des}.
It appears that the overall sign discrepancy is due to 
convention alone,
and so we will detail the interaction
terms that were used in determining the sign of the asymmetry.
Firstly, the strong interaction Lagrange density is 
\begin{eqnarray}
{\cal L}^{st} & = & g_A\ N^\dagger {\bf \sigma}\cdot {\bf A} N
\ =\ {g_A\over f_\pi}\ N^\dagger {\bf \sigma}\cdot {\bf \nabla}\Pi N
\ +\ {\cal O}\left(\Pi^3\right)
\ \ \ ,
\end{eqnarray}
where 
${\bf A} = {i\over 2}
\left( \xi\nabla\xi^\dagger - \xi^\dagger\nabla\xi \right)$
transforms as ${\bf A}\rightarrow U {\bf A} U^\dagger$ under 
$SU(2)_L\otimes SU(2)_R$ chiral transformations.
The $\xi$ field is defined as
\begin{eqnarray}
\Sigma & = & \xi^2 \ =\ 
\exp\left({2 i\over f_\pi}\Pi\right)
\qquad ,\qquad
\Pi  =  
\left(\matrix{\pi^0/\sqrt{2} & \pi^+\cr \pi^- & -\pi^0/\sqrt{2}}\right)
\ \ \ ,
\end{eqnarray}
and under chiral transformations $\Sigma\rightarrow L\Sigma R^\dagger$.
It is straightforward to derive the Noether current associated with 
axial transformations which, when compared to the $\beta$-asymmetry in
$n\rightarrow pe^-\overline{\nu}_e$, fixes $g_A\sim +1.25$
(for instance, see Refs.~\cite{EWbook,Georgi}).
Secondly, the leading order isovector PV interactions are defined by the 
Lagrange density
\begin{eqnarray}
{\cal L}^{wk} & = & 
-{\hpi f_\pi\over 4} N^\dagger \left( X_L^3-X_R^3\right) N
\ =\ 
-{\hpi \over \sqrt{2}} N^\dagger \left[{\bf \tau}, \Pi\right]^3 N
\ +\ {\cal O}\left(\Pi^3\right)
\ \ \ ,
\end{eqnarray}
where $X_L^a = \xi^\dagger\tau^a\xi$, $X_R = \xi\tau^a\xi^\dagger$, transform
as $X_{L,R}\rightarrow UX_{L,R}U^\dagger$ under chiral transformations.
These definitions lead to an opposite sign for the PV pion-exchange potential
~\footnote{
Ref.~\cite{Des} uses a $\gamma_5$ sign convention such that the
left-handed chirality projector is $1+\gamma_5$ 
(e.g. eq.~(29) in Ref.~\cite{DDH})
and 
defines the strong coupling $g_{\pi NN}$ through
$\tilde {\cal L} = 
-i g_{\pi NN} \overline{N}\gamma_5 \pi N$
(eq.~(113) in Ref.~\cite{DDH}).
This requires that $g_{\pi NN}$ is negative for the pion field we have
defined in the text.
Ref.~\cite{Des} assumes that $g_{\pi NN}$ is positive.
} 
compared to that of Ref.~\cite{DDH}.
To reconcile this difference one finds $\hpi$ used in these defintions is of
opposite sign to that used in Ref.~\cite{Des}, despite the fact that the 
weak Lagrange densities appear to be identical.
This set of sign conventions will only matter when there is a rigorous
theoretical prediction for the sign of $\hpi$.
At this point in time no such calculation exists~\footnote{
Absolute signs determined in hadronic models are not necessarily consistent
with QCD.
}.
A comparison with existing hadronic model estimates will provide information 
about the validity of such models for nonleptonic matrix elements~\cite{Des},
and we will have to be satisfied with this somewhat primitive 
level of understanding until a result from lattice-QCD is computed.

In the context of KSW power-counting, where a unified theory of pion-nucleon
and nucleon-nucleon interactions was proposed, the 
estimated $\sim 30\%$ in $A_\gamma$ uncertainty reflects not
only the presence of higher dimension pion-nucleon parity violating
interactions, but higher order contributions to
the strong nucleon-nucleon interaction.
Given the present convergence problems~\cite{FMSa} 
of KSW power-counting 
for certain quantities one cannot read too much into this uncertainty.
However, it remains true that until a converging expansion for the 
NN interaction, consistent with chiral symmetry is constructed, it is 
unrealistic to claim  precision predictions for any observables of this
type within either the potential model approach~\footnote{
In the relevant potential model calculations,  local gauge invariant 
operators involving one-photon and four nucleons
have not been included.
Such operators are certainly present, and may or may not be saturated by
meson-exchange currents.
In the case of the strong width for $np\rightarrow d\gamma$, the pion-exchange
current saturates the local counterterm that appears in $\nopi$~\cite{Ch99}.
In contrast, 
the local counterterm for the deuteron quadrupole moment is {\it not}
saturated by pion-exchange currents~\cite{Ch99}, nor is the 
local counterterm that contributes at NLO to $\nu d$ break up~\cite{BuCh}.
If $\hpi$ is much smaller than NDA estimates, such formally higher
dimension operators will make an enhanced contribution to electromagnetic
observables.}
or using an EFT where 
pions are  included as dynamical degrees of freedom.
At present, claims such as those of Ref.~\cite{Des} are unfounded.


\section{Discussion}

We have shown how to describe the low-energy PV interactions in the 
two-nucleon sector by local operators in $\nopit$.
Given the possibility that $\hpi$ may be much smaller than 
NDA estimates, 
it is fruitful to consider the most general structure of the PV interactions.
This attitude has been taken recently in Ref.~\cite{KKnp} to describe 
the circular polarization asymmetry in $\vec\gamma d\rightarrow np$.
We have shown that there is a model independent relation between $A_\gamma$,
PV asymmetry in  $\vec np\rightarrow d\gamma$, 
and $A_h$, the non-nucleon 
part of the deuteron anapole  moment at leading order in the $\nopit$
expansion.
It is important to test this relation if possible
but it would require the measurement of $A_\gamma$, 
the nucleon anapole moment, and the deuteron anapole moment.

\vskip 0.5in

I would like to thank Silas Beane and Roxanne Springer 
for useful discussions.
This work is supported in part by the U.S. Dept. of Energy under Grants No.
DE-FG03-97ER4014.


\begin{references}

\bibitem{sample} D.T. Spayde {\it et al}.,
(SAMPLE Collaboration),
{\em Phys. Rev. Lett.} {\bf 84}, 1106 (2000).

\bibitem{cesium} S.L. Gilbert, M.C. Noecker, R.N. Watts, C.E. Wieman,
\Journal{\PRL}{55}{2680}{1985};
S.L. Gilbert, C.E. Wieman, \Journal{\PRA}{34}{792}{1986};
M.C. Noecker, B.P. Masterson, C.E. Wieman, \Journal{\PRL}{61}{310}{1988};
C.S. Wood, UMI-97-25806-mc (microfiche), 1997. (Ph.D.Thesis);
C.S. Wood, S.C. Bennet, D. Cho, B.P. Masterson, J.L. Roberts, C.E. Tanner,
and C.E. Wieman, \Journal{\SCI}{275}{1759}{1997};
S.C. Bennett and C.E. Wieman 
{\em Phys. Rev. Lett.} {\bf 82}, 2484 (1999).
 
\bibitem{lightN} C.A. Barnes, M.M. Lowry, J.M. Davidson, 
R.E. Marrs, F.B. Moringo, B. Chang, E.G. Adelberger and  H.E. Swanson
\Journal{\PRL}{40}{840}{1978};
P.G. Bizetti, T.F. Fazzini, P.R. Maurenzig, A. Perego, G. Poggi, P. Sona,
and N. Taccetti,
\Journal{\NCA}{29}{167}{1980};
G. Ahrens, W. Harfst, J.R. Kass, E.V. Mason, H. Schrober, 
G. Steffens, H. Waeffler, P. Bock, and K. Grotz,
\Journal{\NPA}{390}{496}{1982};
S.A. Page, H.C. Evans, G.T. Ewan, S.P. Kwan, J.R. Leslie, 
J.D. MacArthur, W. McLatchie, P.
Skensved, S.S. Wang, H.B. Mak, A.B. McDonald, C.A.
Barnes, T.K. Alexander, E.T.H. Clifford,
\Journal{\PRC}{35}{1119}{1987};
M. Bini, T.F. Fazzini, G. Poggi, and N. Taccetti,
\Journal{\PRL}{55}{795}{1985}.

\bibitem{BSpv}P. F. Bedaque and  M. J. Savage
\Journal{\PRC}{62}{018501}{2000};
J. -W. Chen, T. D. Cohen and C. W. Kao,
{\tt nucl-th/0009031}.

\bibitem{Chpv}J. -W. Chen and Xiangdong Ji,
{\tt hep-ph/0011230};
{\tt nucl-th/0011100}.


\bibitem{DDH}B. Desplanques, J.F. Donoghue and B.R. Holstein, Ann. of
  Phys. {\bf 124}, 449, (1980);
R.D.C. Miller and B.H.J. McKellar, Phys. Reports {\bf 106}, 169
(1984);V.M. Dubovik and S.V. Zenkin, Ann. of Phys. {\bf 172}, 100,
(1986).

\bibitem{twopi}
R. J. Blinn-Stoyle,
{\em Phys. Rev.} {\bf 118}, 1605 (1960);
B. Desplanques,
\Journal{\PLB}{41}{461}{1972};
H. J. Pirner and D. O. Riska,
\Journal{\PLB}{44}{151}{1973};
E. Fischbach and D. Tadic,
{\em Phys. Rep.} {\bf 6C}, 123 (1973);
M. Chemtob and B. Desplanques,
\Journal{\NPB}{78}{139}{1974}.

\bibitem{KS}D.B. Kaplan and M.J. Savage,
\Journal{\NPA}{556}{653}{1993};
\Journal{\NPA}{570}{833}{1994}(E).

\bibitem{ZPHM}
S. -L.  Zhu, S. J. Puglia, B.R. Holstein and  
M.J. Ramsey-Musolf,
{\tt hep-ph/0005281}.


\bibitem{Da65}
G. S. Danilov, 
{\em Phys. Letts.} {\bf 18}, 40 (1965).

\bibitem{Mi76}
J. Missimer,
\Journal{\PRC}{14}{347}{1976}.

\bibitem{DeMi78}
B. Desplanques and J. Missimer,
\Journal{\NPA}{300}{286}{1978}.



\bibitem{Ka97}
D.~B. Kaplan,
\newblock {\em Nucl. Phys.} {\bf B494}, 471 (1997).

\bibitem{BS00}
S. R. Beane and M. J. Savage,
{\tt nucl-th/0011067}.

\bibitem{Ch99}
J.-W. Chen, G.~Rupak, and M.~J. Savage,
\newblock {\em Nucl. Phys.} {\bf A653}, 386 (1999).


\bibitem{KKnp}
I. B. Khriplovich and R. V. Korkin,
{\tt nucl-th/0010032}.

\bibitem{snow}M. Snow {\it et al}., 
{\em Nucl. Inst. and Meth.} {\bf 440}, 729 (2000).

\bibitem{KSSWa} D. B. Kaplan, M. J. Savage, R. P. Springer 
and M. B. Wise,
{\em Phys. Lett.} {\bf B449}, 1 (1999).

\bibitem{LaMC76}
K. R. Lasset and B. H. J. McKellar,
\Journal{\NPA}{260}{413}{1976}.

\bibitem{SSa} M. J. Savage and R. P. Springer,
{\em Nucl. Phys.} {\bf A644}, 235 (1998);
{\em Nucl. Phys.} {\bf A657}, 457 (1999). 

\bibitem{SSb} M. J. Savage and R. P. Springer,
{\tt nucl-th/9907069}.
 
\bibitem{HolMus}W.C. Haxton, E.M. Henley and M.J. Musolf,
\Journal{\PRL}{63}{949}{1989}.


\bibitem{vKM}C.M. Maekawa, J.S. Veiga and U. van Kolck ,
{\em Phys. Lett.} {\bf B488}, 167 (2000);
 C.M. Maekawa and U. van Kolck,
{\em Phys. Lett.} {\bf B478}, 73 (2000).
 
\bibitem{HeHw82}
E. M. Henley and W-Y. P. Hwang,
\Journal{\PRC}{26}{2376}{1982}.

\bibitem{KKana} 
I.B. Khriplovich and R.A. Korkin
{\em Nucl. Phys.} {\bf A665}, 365 (2000 ).


\bibitem{Ka98}
D.~B. Kaplan, M.~J. Savage, and M.~B. Wise,
\newblock {\em Nucl. Phys.} {\bf B534}, 329 (1998);
\newblock {\em Phys. Lett.} {\bf B424}, 390 (1998);
\newblock {\em Phys. Rev.} {\bf C59}, 617 (1999).


\bibitem{Des} B. Desplanques,
{\tt nucl-th/0006065}.

\bibitem{EWbook} 
Chapter 9 of 
{\it Pions and Nuclei}, by
T. Ericson and W. Weise,
Oxford Science Publications (1988).
ISBN 0-19-852008-5.

\bibitem{Georgi} 
{\it Weak Interactions and Modern Particle Theory},
by H. Georgi,
Addison-Wesley Publishing Company (1984).
ISBN 0-8053-3163-8.

\bibitem{FMSa}
S. Fleming, T. Mehen and I. W. Stewart,
\Journal{\NPA}{677}{313}{2000}.


\bibitem{BuCh}
M.N. Butler and J.-W. Chen,
\Journal{\NPA}{675}{575}{2000};
\Journal{\PRC}{63}{035501}{2001};
{\tt nucl-th/0101017}.

\end{references}
\end{document}